\documentclass[11pt,a4paper]{article}
\usepackage{jcappub}
\usepackage{latexsym,hyperref}
\usepackage{jcappub}

\usepackage{graphicx}	
\usepackage{amsmath}	
\usepackage{pdflscape}
\usepackage{xcolor}
\usepackage{ulem}

\newcommand{\beq}{\begin{equation}}
\newcommand{\eeq}{\end{equation}}
\newcommand{\bea}{\begin{eqnarray}}
\newcommand{\eea}{\end{eqnarray}}
\def\be{\begin{equation}}
\def\ee{\end{equation}}
\def\ba{\begin{eqnarray}}
\def\ea{\end{eqnarray}}

\def\nn{\nonumber}
\newcommand{\bi}{\begin{itemize}}
\newcommand{\ei}{\end{itemize}}
\newcommand{\efi}{\end{figure}}
\newcommand{\efib}{\end{figure}}
\newcommand{\efig}{\end{figure*}}

\newcommand{\hompc}{\,h\,{\rm Mpc}^{-1}}

\newcommand{\mpcoh}{\,h^{-1}\,{\rm Mpc}}

\title{Model-independent parameters from DESI DR1 LRG2 with an explicitly varying power-spectrum shape}
\author[a]{Yong-Seon Song,}
\author[b,c]{M. Icaza-Lizaola,}
\author[d]{Minji Oh}
\author[e,f]{and Yi Zheng}

\affiliation[a]{Korea Astronomy and Space Science Institute,  776 Daedeok-daero, Yuseong-gu, Daejeon 34055, Republic of Korea}
\affiliation[b]{Departamento de F\'isica Te\'orica,  Universidad Aut\'onoma de Madrid, 28049 Madrid, Spain.}
\affiliation[c]{Centro de Investigación Avanzada en Física Fundamental (CIAFF), Universidad Autónoma de Madrid, ES-28049 Madrid, Spain.}
\affiliation[d]{Chosun University, 375 Seosuk-dong, Dong-gu, Gwangjiu 501-759, Republic of Korea.}
\affiliation[e]{School of Physics and Astronomy, Sun Yat-sen University, 2 Daxue Road, Tangjia, Zhuhai, 519082, China.}
\affiliation[f]{CSST Science centre for the Guangdong-Hong kong-Macau Greater Bay Area, SYSU, Zhuhai, 519082, China.}

\emailAdd{ysong@kasi.re.kr}
\emailAdd{miguel.deicaza@uam.es}
\emailAdd{minji.wow@gmail.com}
\emailAdd{zhengyi27@mail.sysu.edu.cn}

\abstract{

Distinguishing between dark energy and modified gravity as the origin of the accelerated expansion requires the cosmic distances and the growth of structure to be measured in a model-independent way. Such measurements are difficult to perform while the primordial shape parameters are varied at the same time, and those parameters are therefore usually held fixed at the values determined by cosmic microwave background experiments. In this work, we present a methodology that constrains the shape parameters simultaneously with the model-independent distance and growth parameters, using an emulator built upon a hybrid model in which perturbative predictions are supplemented by templates measured from $N$-body simulations. {Because these templates are measured rather than expanded, the model carries non-linear corrections beyond the order at which its perturbative part is truncated, so that its accuracy on small scales is set by the simulations rather than by the reach of the expansion.} A Gaussian process emulator supplies these templates rapidly across the space of scale-dependent parameters, which makes the joint analysis computationally feasible, and the emulator is trained and validated over a range spanning $5\sigma$ of the Planck constraints. Applying the method to the LRG2 sample of DESI DR1, we obtain measurements of the Alcock--Paczynski parameters $\alpha_\perp$ and $\alpha_{||}$ and of the growth rate $f\sigma_8$. {With the tight Planck prior adopted here the constraints differ from those obtained with the shape parameters held fixed by less than a twentieth of their own uncertainties, so that the choice of template contributes at that level to the error budget of the measurement. Projected onto flat $\Lambda$CDM after the fit, the three measurements are mutually consistent, with $\chi^2=0.7$ for two degrees of freedom, and require $h=0.724\pm0.017$; the preference for an isotropic distance smaller than that of the fiducial cosmology is the one DESI reports for this sample from both its full-shape and its baryon-acoustic-oscillation analyses.} The value of the approach lies in what it enables next: once the Planck prior is relaxed, the same framework can test directly whether the shape parameters preferred by galaxy clustering are consistent with those inferred from the cosmic microwave background. 
}

\begin{document}
\maketitle
\flushbottom

\section{Introduction}
\label{sec:intro}

Redshift surveys of galaxies have played a key role in probing the large-scale structure of the Universe over the past decades {\citep{Colless:2001gk,York:2000gk,Alam:2016hwk,Alam:2020sor}}. The three-dimensional distribution of dark matter fluctuations across cosmological volumes can be traced through precise measurements of galaxy clustering, obtained from the redshifts of vast numbers of galaxies. This spatial distribution and the peculiar motions of galaxies {\citep{Kaiser:1987qv}} encode key information about the Universe, including its initial conditions, the expansion history, the growth of cosmic structure, and the nature of gravity. Measuring galaxy clustering deep into the Universe therefore provides an opportunity to test a wide range of theoretical models for both the initial conditions of the Universe and the origin of its late-time cosmic acceleration.

Two decades after the discovery of cosmic acceleration {\citep{Riess:1998cb,Perlmutter:1998np}}, its physical cause remains unresolved. Many theoretical models have been proposed, and they broadly fall into two classes. The first invokes dark energy, an unknown energy component added to the cosmic budget {\citep{Copeland:2006wr}}; the second invokes modified gravity, in which the gravitational physics of Einstein’s relativity is altered on cosmological scales {\citep{Clifton:2011jh}}. Although cosmic distances are now measured precisely by multiple experiments {\citep{2020A&A...641A...6P,Alam:2020sor}}, the two classes cannot generally be distinguished by distance measurements alone. Once the growth of structure is additionally observed, however, the effects of modified gravity, including those associated with screening mechanisms {\citep{Vainshtein:1972sx,Khoury:2003rn}}, can also be probed {\citep{Linder:2005in,Guzzo:2008ac}}. Combining measurements of cosmic distances and structure growth therefore provides an important means of distinguishing between these different explanations for cosmic acceleration. In this paper we therefore advocate for a methodology in which both cosmic distances and growth factors are measured in a model-independent way {\citep{Song:2008qt,2021PhRvD.104d3528S}}, so that a wide range of dark energy and modified gravity models can be tested. {The shape of the primordial power spectrum is the one ingredient such an analysis cannot leave free without a model of its own, and it is usually held at the values measured by cosmic microwave background experiments \citep{2020A&A...641A...6P}, or else compressed into an effective shape parameter added to the fit \citep{Brieden:2021edu}.}

The Dark Energy Spectroscopic Instrument (DESI) {\citep{DESI:2016fyo,Adame_2024}} is the most ambitious spectroscopic galaxy survey to date, aiming to collect spectra for over 35 million galaxies and quasars across 14,000 square degrees of sky. One of its primary scientific goals is to probe the expansion history and the growth of structure in the Universe through measurements of Baryon Acoustic Oscillations (BAO) and Redshift-Space Distortions (RSD). DESI Data Release 1 (DR1) is now available \citep{2025JCAP...09..008A}{, with the large-scale structure catalogues and the two-point clustering measurements described separately \citep{Ross2024,DESI2024II}}, and in this work we focus on the Luminous Red Galaxy sample in the redshift range \(0.6<z<0.8\) (LRG2). In~\cite{2025JCAP...09..008A}, the fitted AP parameters of this sample showed the largest deviations from the Planck fiducial values. While the original DESI DR1 full-shape analysis employed an EFT-based model {\citep{Ivanov:2019pdj,DAmico:2019fhj}}, here we analyse the same sample using an alternative approach based on our hybrid model.

In this work we focus on RSD in the large-scale structure as a promising probe of both cosmic expansion and the growth of structure simultaneously. The anisotropic features of the power spectrum along the line of sight are induced by the peculiar motions of galaxies {\citep{Kaiser:1987qv}}, and a careful analysis of these features yields information about the growth of structure. The same measurement also constrains the cosmic expansion history precisely, through the BAO method and the Alcock--Paczynski (AP) test {\citep{AP_1979Natur.281..358A}}. A single RSD measurement therefore provides access to two key cosmological observables.

An emulator method {\citep{2024MNRAS.535..588I}} based upon the hybrid model {\citep{2016JCAP...08..050Z,2018JCAP...07..018S}} is adopted to compute accurate theoretical power spectra up to $k\le 0.2\hompc$. The cosmological parameters are categorized into two distinct groups, namely scale-dependent and \textit{scale-independent} parameters. The \textit{scale-dependent} parameters determine the shape of the power spectrum in the primordial epoch, which is set by the initial conditions and by the competition between gravity and radiation pressure; the emulated templates for this shape are generated from simulations combined with a theoretical model that goes beyond second-order perturbation theory {\citep{2012PhRvD..86j3528T,PhysRevD.89.023502}}. The \textit{scale-independent} parameters, in turn, describe the coherent evolution of the power spectrum in the linear regime at later times, so that the growth through the epoch of cosmic acceleration is predicted on top of the shape fixed by the emulator. Following \cite{ZhengY2019,2021PhRvD.104d3528S}, we further extend the model to account for galaxy bias, which allows the galaxy power spectrum to be constructed from the matter power spectrum. The emulated theoretical power spectrum is verified to reproduce the measured power spectrum accurately, and since the model incorporates perturbative corrections beyond second order, making the framework suitable for application to surveys such as DESI.

The paper is organized as follows. In Section~\ref{sec:HybridModel}, we introduce our galaxy RSD mapping formula and the galaxy density and velocity bias models. In Section~\ref{sec:mock_analysis}, we validate the methodology against mock galaxy catalogues and assess its accuracy. In Section~\ref{sec:data_analysis}, we apply the methodology to the DESI DR1 data. Finally, the conclusions and discussion are presented in Section~\ref{sec:Conclucion}.

\section{The Hybrid model}
\label{sec:HybridModel}

Throughout this chapter, we present our hybrid modeling methodology for computing matter power spectrum templates that extends beyond two-loop perturbation theory (PT) by incorporating higher-order statistics. The hybrid model was originally introduced in \cite{2016JCAP...08..050Z,2018JCAP...07..018S}, where a set of scaling relations was used to capture the effects of \textit{scale-independent} parameters on the power spectrum. This approach was later refined by \cite{2024MNRAS.535..588I}, where emulators are employed to model the impact of scale-dependent parameter variations. These emulators are trained to reproduce target statistics with high fidelity, closely matching results from full N-body simulations. Since such simulations are computationally expensive to run, emulators offer an efficient alternative. In \cite{2024MNRAS.535..588I}, it was shown that the emulators reproduce the relevant statistics with an accuracy better than 0.1 per cent within the $5\sigma$ confidence region of the Planck 2018 cosmology.

The whole methodology is fairly complex and is thoroughly discussed in the works we just cited, in here we include a brief summary of the method and we refer the reader to the original paper for a detailed and thorough discussion and validation of the method.

Our hybrid framework models the redshift-space power spectrum as a function of both \textit{scale-independent} and \textit{scale-dependent} cosmological parameters, as well as a small set of nuisance and geometric parameters. The \textit{scale-dependent parameters} are those that affect the shape of the power spectrum and include the cold dark matter density \( \omega_{\rm c} = \Omega_{\rm c} h^2 \), the baryon density \( \omega_b = \Omega_b h^2 \), and the scalar spectral index \( n_s \). These parameters are explored using N-body simulations and emulated through Gaussian Process models. The \textit{scale-independent parameters} primarily control the amplitude and overall normalization of the power spectrum. Their effect is captured analytically via scaling relations that act as a higher order correction on perturbation theory predictions, this scaling relations depend on two growth parameters \( G_\delta = D_+ \) and \( G_\theta = f D_+ \), which parameterize the linear growth of density and velocity fields. In addition, our model includes a velocity dispersion nuisance parameter \( \sigma_p \), which regulates the small-scale damping via a Finger-of-God term, and two Alcock–Paczynski parameters \cite{AP_1979Natur.281..358A}, \( \alpha_\parallel \) and \( \alpha_\perp \), which account for distortions due to differences between the fiducial and true cosmologies when mapping redshifts to distances.

Following \cite{PhysRevD.82.063522,2016JCAP...08..050Z} we can express the matter power spectrum as the product of a perturbative and a non-perturbative contribution as
\begin{align}
  P_s(k,\mu) &= D^{\mathrm{FoG}}\!\bigl(k\,\mu\,\sigma_p\bigr)\,
  \Bigl\{
        P_{\delta\delta}(k) + 2\mu^2 P_{\delta\theta}(k) + \mu^4 P_{\theta\theta}(k)
        \nonumber\\[2pt]
        &\qquad
        +\,A(k,\mu) + B(k,\mu) + F(k,\mu) + T(k,\mu)
  \Bigr\}.
  \label{eq:eq8}
\end{align}
The terms \(P_{\delta\delta}\), \(P_{\Theta\Theta}\), and \(P_{\delta\Theta}\) represent the power spectra associated with the density field, the velocity divergence field (\(\Theta = \nabla \cdot \vec{u}\)), and their cross-correlation, respectively. To account for non-perturbative effects, particularly those arising from small-scale random peculiar motions -- commonly referred to as the Finger-of-God (FoG) effect -- we include a phenomenological damping factor \(D^{\rm FoG}(k,\mu,\sigma_p)\); the specific functional form adopted for this term is a fitting-methodology choice, which we describe later in this work. The remaining corrections, denoted by \(A\), \(B\), \(F\), and \(T\), encapsulate higher-order contributions beyond the leading perturbative terms and are crucial for accurately modeling the power spectrum on quasi-linear scales.

Our Gaussian Process (GP) emulators are trained to learn the mapping  
\[
\{\omega_{\mathrm{c}}, \omega_b, n_s\} \mapsto \{\bar{P}_{\delta\delta}, \bar{P}_{\delta\theta}, \bar{P}_{\theta\theta}, \bar{A}, \bar{B}, \bar{F}, \bar{T}\},
\]  
where the overbar denotes results computed with the \textit{scale-independent} parameters fixed to their fiducial values. Once the emulator provides predictions for these statistics at a given point in parameter space, we must still incorporate the contributions from the \textit{scale-independent} parameters into the final model. Below, we briefly summarize how this correction is applied for the terms \( \bar{P}_{XY} \), corresponding to the fields \( X, Y \in \{\delta, \Theta\} \). The corrections for the \(\{\bar{A}, \bar{B}, \bar{F}, \bar{T}\}\) terms are not discussed here but can be found in the appendix of \cite{2024MNRAS.535..588I}.

First, we compute a baseline prediction using two-loop perturbation theory (PT), specifically following the \textsc{RegPT} formalism of \cite{2012PhRvD..86j3528T}. Although \textsc{RegPT} calculations are relatively fast when using public codes, we further accelerate this step by building a second set of GP emulators trained on the \textsc{RegPT} outputs, as described in \cite{2024MNRAS.535..588I}. This baseline prediction allows us to compute a residual correction term, defined as the difference between the N-body-based prediction (from the first emulator) and the corresponding PT prediction. This residual captures the higher-order nonlinear contributions not included in the two-loop PT model.

Specifically, for a given set of scale-dependent parameters, we compute the residual:
\begin{equation}
  \bar{P}^{\rm res}_{XY}(k, z) = \bar{P}^{\rm PT}_{XY}(k, z) - \bar{P}^{\rm N\text{-}body}_{XY}(k, z),
  \label{eq:residual}
\end{equation}
where \( \bar{P}^{\rm PT}_{XY} \) denotes the two-loop PT prediction computed using the \textsc{RegPT} formalism, and \( \bar{P}^{\rm N\text{-}body}_{XY} \) is the emulator prediction based on N-body simulations. Both quantities are evaluated at the same values of the scale-dependent parameters, while the \textit{scale-independent} parameters are fixed to their fiducial values during this step.

Once the residual is obtained, we apply a scaling relation to translate it to a target cosmology with different \textit{scale-independent} parameters:
\begin{equation}
  P^{\rm res}_{XY}(k, z) =
  \frac{G_X G_Y G_\delta^4}{\bar{G}_X \bar{G}_Y \bar{G}_\delta^4}
  \bar P^{\rm res}_{XY}(k, z),
  \label{eq:scaling}
\end{equation}
where \(G_X\) and \(G_Y\) (with \(X, Y \in \{\delta, \theta\}\)) are the growth functions evaluated in the target cosmology, and the barred quantities refer to the fiducial cosmology used to generate the emulator predictions. The final prediction for the power spectrum is then obtained by adding the scaled residual to the PT prediction evaluated in the target cosmology.

\section{Model-independent data analysis}
\label{sec:mock_analysis}

\begin{figure}
\begin{center}
\includegraphics[width=0.710\textwidth]{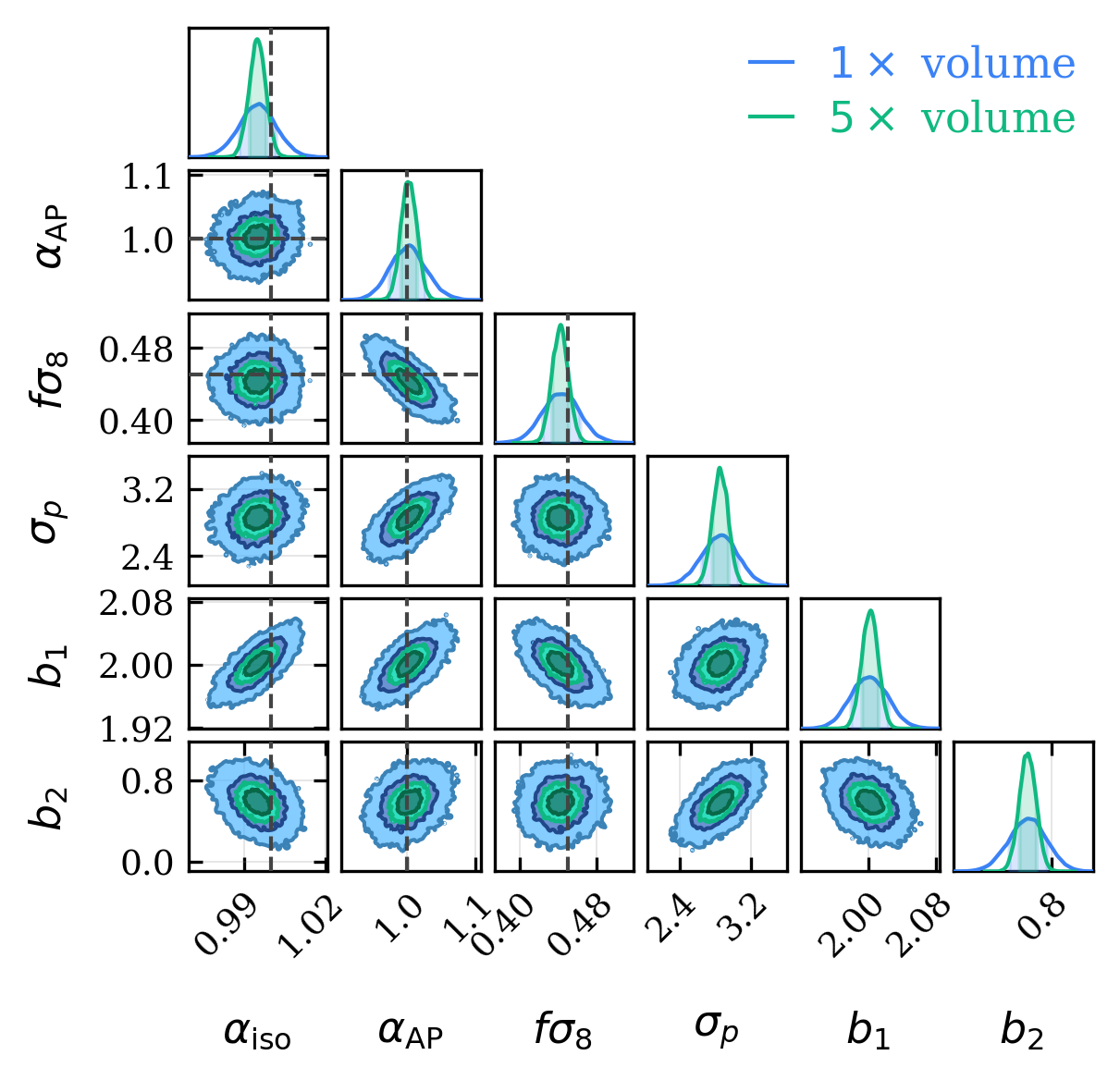}
\end{center}
\caption{{The corner plot for the six parameters $(\alpha_{\rm iso},\alpha_{\rm AP},f\sigma_8,\sigma_p,b_1,b_2)$ recovered from the single-template analysis at $z=0.8$, showing the inner 68 and outer 95 per cent confidence regions. The blue contours are measured from the nominal DESI survey volume and the green contours from a volume five times larger; overlaying the two on common axes makes the gain in precision directly readable. The dilation parameters are given in the DESI convention, $\alpha_{\rm iso}=(\alpha_{||}\alpha_\perp^2)^{1/3}$ and $\alpha_{\rm AP}=\alpha_{||}/\alpha_\perp$, and are formed sample by sample from the chain. Black dashed lines mark the input values of the simulation, $\alpha_{\rm iso}=\alpha_{\rm AP}=1$ and $f\sigma_8=0.4506$. The $f\sigma_8$ axis carries the physical velocity amplitude rather than the raw chain parameter, which the fit returns as $f\sigma_8\sqrt{A_s/A_s^*}$; the two differ by a constant factor and only the former can be compared with the input value. the nuisance parameters $\sigma_p$, $b_1$ and $b_2$ have no fiducial value assigned. $\sigma_p$ is given in units of $\mpcoh$. The figure is drawn at the same scale as Fig.~\ref{fig:multi_corner_sim}, so that a cell of one corner plot is the same size as a cell of the other.}} 
\label{fig:single_corner_sim}
\end{figure}

\subsection{Abacus simulation}
\label{subsec:abacus}
\label{subsec:simulations}

Prior to analyzing the DESI DR1 sample, we rigorously validate our hybrid RSD methodology 
and emulator performance 
against the \textsc{AbacusSummit}\footnote{https://abacussummit.readthedocs.io/en/latest/cosmologies.html} 
cosmological $N$-body simulation suite \citep{10.1093/mnras/stab2484}, which consists of a massive set of high-accuracy, high-resolution $N$-body simulations, produced with the Abacus $N$-body code \citep{10.1093/mnras/stab2482}. As a standard benchmark within the DESI community, it provides a high-fidelity mock dataset for the purpose of validating data analysis pipeline. For this verification, we restrict our analysis to the \textit{Planck} 2018 $\Lambda$CDM cosmology \citep{2020A&A...641A...6P}, designated as \textsc{abacus\_cosm000}. The precise cosmological parameters defining this fiducial universe are fixed as follows: baryon density $\omega_b \equiv \Omega_b h^2 = 0.02237$, cold dark matter density $\omega_{\rm c} \equiv \Omega_c h^2 = 0.1200$, Hubble parameter $h=0.6736$, primordial spectral index $n_s = 0.9649$, and the amplitude of primordial scalar fluctuations $A_s = 2.0830 \times 10^{-9}$ evaluated at a pivot scale of $k_{0} = 0.05 \rm{Mpc}^{-1}$. Consistent with the baseline Planck configuration, the simulation incorporates a massive neutrino component with $\Omega_{\nu}h^2 = 0.0006442$, providing a total neutrino mass sum of $\sum m_{\nu} =0.06 \, \rm{eV}$.

We utilize the standard ``base'' realization of \textsc{AbacusSummit}, which models the dark matter distribution inside a periodic comoving cubic volume of side length $L_{\rm box} = 2\,h^{-1}\rm{Gpc}$ with $N_p = 6912^3$ particles. This configuration yields a high mass resolution of $m_p = 2 \times 10^9 h^{-1}M_{\odot}$, ensuring enough convergence down to the small scales of our interest targeted by our emulator. Halos are identified on-the-fly using a method called \textsc{COMPASO} \citep{10.1093/mnras/stab2980}. Galaxy catalogues are generated using the extended halo occupation distribution (HOD) model implemented in the \textsc{AbacusHOD} code for the dark time tracers including LRG2 \citep{10.1093/mnras/stab3355}. The HOD parameters are found by fitting against the two-point correlation measurements of the DESI early data release \citep{Adame_2024}. Details on the mock generation process are described in Section 3 of \cite{2025JCAP...09..008A}. To construct our validation sample, we extract the redshift-space clustering statistics using a geometric cutsky mock generated from the snapshot at $z=0.8$, strategically chosen to match the effective redshift range of the DESI DR1 LRG2 sample ($0.6 < z < 0.8$). In analogy with our previous study, the matter density fields in redshift space are computed from the mock catalogs using a standard Cloud-in-Cell (CIC) mass assignment scheme, followed by corrections for the grid interlacing effects and shot noise. This mock data verification enables a robust test on the systematics of our hybrid RSD approach before evaluating real data.

\subsection{Verification test for Single-template data analysis}\label{subsec:single_verify}

\begin{figure}
\begin{center}
\includegraphics[width=\textwidth]{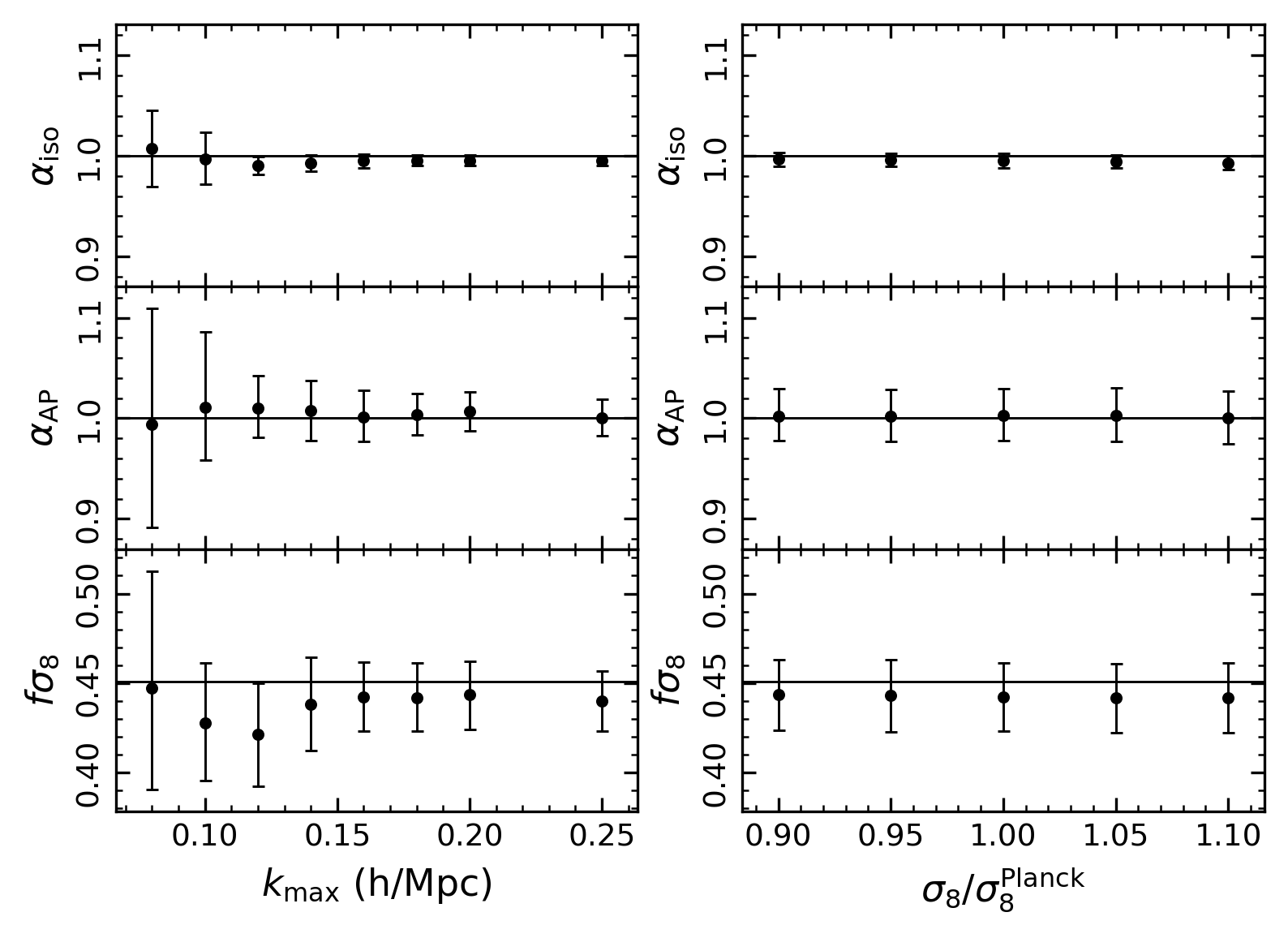}
\end{center}
\caption{{\it (Left panel)}: The constraints on the three key cosmological parameters $(\alpha_{\rm iso},\alpha_{\rm AP},f\sigma_8)$ from the single-template analysis, shown as a function of the maximum wavenumber $k_{\rm max}$ included in the fit. The horizontal lines mark the fiducial values, and the error bars show the 68 per cent confidence intervals. {\it (Right panel)}: The same constraints shown as a function of the assumed $\sigma_8$, normalized to its fiducial value.} 
\label{fig:cos_constraints_sim_varying}
\end{figure}

{To validate our single-template methodology, we fit the monopole and quadrupole of the redshift-space galaxy power spectrum measured from the AbacusSummit LRG2 mock catalogues described in Section \ref{subsec:simulations}. These measurements provide a test of our ability to recover the input cosmology from galaxy clustering using the hybrid RSD model. In this analysis, the shape of the matter power spectrum is kept fixed, while the cosmological and nuisance parameters are varied simultaneously.}

In a single-template analysis of the anisotropic power spectrum, we aim to probe the cosmological parameters in a way that is independent of the dark energy model. The target parameters are the growth factor $f\sigma_8$ and the cosmic distances $D_A$ and $H^{-1}$, which are expressed by the AP parameters $\alpha_\perp$ and $\alpha_{||}$, respectively. These key cosmological parameters are marginalized over the nuisance parameters, namely the FoG parameter $\sigma_p$ and the local bias parameters $b_1$ and $b_2$. The full parameter space in the single-template case is therefore the six-dimensional set $(\alpha_\perp,\alpha_{||},f\sigma_8,\sigma_p,b_1,b_2)$.

An accurate probe of the growth function $f\sigma_8$ is mainly contaminated by non-linear corrections arising from the random motions of galaxies, which damp the anisotropic power spectrum. This non-perturbative FoG effect cannot be formulated from first principles, and we therefore adopt a simple Gaussian form for the FoG function in this paper, given the lack of a statistically significant difference among typical forms as tested in \cite{Zheng:2018ljz}. In this approximate treatment, the functional form is fixed at the leading-order contribution, and the one-point velocity dispersion $\sigma_p$ is introduced as a free parameter to be marginalized over. {Explicitly, $D^{\rm FoG}(k\mu\sigma_p)=\exp[-(k\mu\sigma_p)^2]$, with $k$ in $\hompc$ and $\sigma_p$ in $\mpcoh$ throughout.} The measured $f\sigma_8$ is then unbiased up to the scales where the leading-order FoG contribution remains dominant, which in turn sets the limitation of our methodology beyond a threshold scale. 

Galaxies are biased tracers of the underlying dark matter field, through which the large-scale structure of the Universe is probed. The relation between the dark matter distribution and the observed galaxy fluctuations is parameterized in this paper by the local linear and non-linear bias parameters, $b_1$ and $b_2$. The non-local effects are described in terms of the leading-order bias parameter by exploiting the consistency relation \citep{Baldauf:2012hs, 1994ApJ...433....1B, 2014PhRvD..90l3522S,PhysRevD.85.083509}. Although this is an approximate description of the full complexity of perturbative galaxy-bias modelling, the simplification remains useful up to mildly non-linear scales, where the non-local contribution to the galaxy mapping is subdominant compared with the local bias effect. The approximate expression for the biased power spectrum $P_{\delta_g\delta_g}$ is given by,
\bea
P_{\delta_g\delta_g}&=&P^2_{\delta_g\delta}/P_{\delta\delta} \,.
\label{eq:pdd_bias}
\eea
where the cross-spectrum between galaxies and dark matter, expanded up to one-loop order, is written as,
\bea
\label{eq:phm_bias}
P_{\delta_g\delta}(k)&=&b_1P_{\delta\delta}(k)+b_2P_{b2,\delta}(k)-\frac{4}{7}(b_1-1) P_{bs2,\delta}(k)
\nonumber \\
&+&\frac{32}{315}(b_1-1)\sigma_3^2(k)P^{\rm{L}}_{\rm m}(k)\,.
\eea
where $P^{\rm{L}}_{\rm m}$ represents the linear matter power spectrum, and $P_{b2,\delta}$, $P_{bs2,\delta}$ and $\sigma_3^2$ are the bias kernels given in \cite{2009JCAP...08..020M}.

Fossil evidence of the primordial baryon-photon plasma, produced by the competition between gravity and radiation pressure, is imprinted on the large-scale structure of the Universe at the last scattering surface. The power spectrum shaped at this primeval epoch then evolves coherently to the present time. The amplitude of the primordial power spectrum $A_S^2$ is provided as prior information from CMB experiments, while the growth of the dark matter density field at large scales is parameterized by $\sigma_8$. When a strong prior on the primordial shape of the power spectrum from CMB experiments is additionally imposed, $\sigma_8$ becomes precisely measurable; we refer to this as a  single-template treatment in this work. Since dark matter is not directly observable, however, large-scale structure formation is traced by galaxies, which introduces a tight degeneracy between $\sigma_8$ and the leading-order galaxy mapping parameter $b_1$. This degenerate combination of the growth of dark matter and the coherent galaxy bias is fully marginalized over by letting $b_1$ float freely while $\sigma_8$ is fixed at its fiducial value. The effect of varying $\sigma_8$ is presented in the right panel of Fig.~\ref{fig:cos_constraints_sim_varying}, which shows that the three key cosmological parameters are insensitive to the variation of $\sigma_8$; only the leading-order bias $b_1$ is anti-correlated with it.

The corner plot of these cosmological and nuisance parameters is presented in Fig.~\ref{fig:single_corner_sim}. While the theoretical RSD model accurately describes the measured spectrum at linear scales, the unknown small-scale physics involved in the large-scale anisotropic clustering becomes non-negligible as the quasi-linear regime is approached. For our fitting methodology, we adopt a Gaussian form to describe this non-perturbative FoG damping; If left unmodeled beyond an appropriate threshold, it would bias the recovered cosmological parameters. The influence of this non-perturbative damping is captured by the single parameter $\sigma_p$ up to an appropriate $k_{\rm max}$. The $(3,3)$ component of Fig.~\ref{fig:single_corner_sim} shows the projected two-dimensional contours between $f\sigma_8$ and $\sigma_p$. The contour is broad in both directions rather than strongly tilted, indicating that $\sigma_p$ is only loosely constrained by the range of scales used here, which is restricted to the mildly non-linear regime below the threshold limit. Our incomplete knowledge of the exact FoG effect therefore degrades the constraint on $f\sigma_8$, although the fiducial value is still successfully recovered. The dependence of the constraints on the choice of $k_{\rm max}$ is presented in the left panel of Fig.~\ref{fig:cos_constraints_sim_varying}, and hereafter $k_{\rm max}$ is fixed to $0.16\hompc$.

The $(3,2)$ and $(3,1)$ components of Fig.~\ref{fig:single_corner_sim} show the contours between the local bias parameters and $f\sigma_8$. The $(3,2)$ corner plot presents the correlation between the linear bias parameter $b_1$ and $f\sigma_8$. Since the density growth function is fixed in this single-template treatment, $b_1$ is constrained with high precision. Although the anti-correlation between the two parameters is significant, the precise determination of $b_1$ does not substantially affect the precision of $f\sigma_8$. The $(3,1)$ corner plot shows the correlation between the next-order local bias parameter $b_2$ and $f\sigma_8$. Below the cutoff scale of $k_{\rm max}=0.16\hompc$, the contribution of $b_2$ is less significant than that of $b_1$, so that $b_2$ is more weakly constrained and its impact on $f\sigma_8$ remains relatively marginal. We note that the contribution of the non-local bias to the measured power spectrum at $k_{\rm max}<0.16\hompc$ is much smaller than that of the second-order local bias parameter, so that no significant bias is induced on $f\sigma_8$; this justifies our use of the consistency relations.

Fig.~\ref{fig:single_corner_sim} presents the same cosmological constraints for the nominal DESI volume and for a volume five times larger, overlaid on common axes. Although mild shifts at the $1\sigma$ level are seen for the three key cosmological parameters, the fiducial values given by the simulations are still successfully recovered.

{It is worth setting out what this test does and does not establish, since the answer depends on the volume at which it is run. At the nominal survey volume we recover $\alpha_{\rm iso}=0.9950\pm0.0068$, $\alpha_{\rm AP}=1.0014\pm0.0266$ and $f\sigma_8=0.4435\pm0.0194$ against an input $f\sigma_8=0.4506$; the isotropic dilation is low by $0.73\sigma$ and the growth rate by $0.37\sigma$, so the input cosmology is recovered without significant bias. The velocity amplitude quoted here is the physical one: what the chain carries is $f\sigma_8\sqrt{A_s/A_s^*}$, and comparing that raw parameter with the template's own $f\sigma_8$ rather than with the input of the simulation is a different quantity and a larger apparent deficit. At five times the volume the error bars shrink to $\pm0.0030$ on $\alpha_{\rm iso}$ and the same offset becomes a $1.7\sigma$ shift. These are the two volumes at which the DESI full-shape analysis validates its own pipeline, and on that criterion ours passes.

The offset itself, however, is real rather than statistical, and we prefer to say so. The error bars scale as the inverse square root of the volume to the digit shown, while the central value does not move at all: $\alpha_{\rm iso}=0.99505$, $0.99500$ and $0.99499$ at one, five and twenty-five times the volume. Against the covariance of the mean of the 25 realisations the same half-per-cent offset is therefore a $3.7\sigma$ statement, and it is a property of the model, not of the realisations. Two things should be said about it. First, its size: half a per cent is a third of the statistical uncertainty on $\alpha_{\rm iso}$ in the DESI DR1 LRG2 measurement of Section~\ref{subsec:DESI_single}, and its sign is such that it cannot account for the low isotropic dilation found there, being a tenth of that shift and in the direction of making it smaller still. It does not affect the measurements reported in this paper. Second, its origin: a residual at this level is at the limit of what the present construction controls, and identifying it is a question for the next revision of the emulator. We would add that half a per cent is also the level at which the question stops being only about the non-linear model. A fit that claims this accuracy rests on a linear power spectrum computed by a Boltzmann code, and whether such a calculation is itself reliable to half a per cent is a prior question that has to be answered before a residual of that size can be attributed to the quasi-linear modelling.}

\begin{table}
\renewcommand{\arraystretch}{1.45} 
\setlength{\tabcolsep}{0.2cm}
\begin{center}
\begin{tabular}{|l|c|c|c|c|}
\hline
\hline
Parameter & Emulator & Emulator' & Emulator'' & Abacus Summit \\
\hline
\hline
$h$       & $0.6727$ & $0.6727$ & $0.6727$ & $0.6736$  \\
\hline
$N_{\nu_0}$       & $3.046$ & $2.0328$ & $2.0328$ & $2.0328$  \\
\hline
$N_{\nu}$       & $0$ & $1$ & $1$ & $1$  \\
\hline
$10^4 \omega_{\nu}$       & $0$ & $6.4420$ & $6.4420$ & $6.4420$  \\
\hline
$10^9 A_S^2$       & $2.1010$ & $2.1010$ & $2.0830$ & $2.0830$  \\
\hline
\hline
$\sigma_8$       & $0.822089462$ & $0.81135$ & $0.8076522$ & $0.807952$  \\
\hline
\hline
\end{tabular}
\end{center}
\caption{The \textit{scale-independent} parameters are presented in this table. The \textit{scale-dependent} parameters are fixed with the following values $\omega_c=0.1202$, $\omega_b=0.02236$ and $n_S=0.9649$ for all cases.} 
\label{tab:cosmopara}
\end{table}

 The emulator is prepared with massless neutrinos, while the test mock catalogue is generated with one massive neutrino. {The two cosmologies therefore differ, as the template and the data always do in an analysis of real observations, and the verification is run without removing that difference. Its effect on the measurement is a normalisation: with a different neutrino mass the amplitude of the linear spectrum at last scattering is not the emulator value, and the measured $f\sigma_8$ is corrected for that shift as set out below.  The residual offsets found in the verification test are not attributed to the neutrino treatment or to the difference in the fiducial cosmologies, and whether those differences contribute to them has not been established.} The scale-dependent parameters are fixed at $\omega_c=0.1202$, $\omega_b=0.02236$ and $n_S=0.9649$, the values used for the template mock. The full list of \textit{scale-independent} parameters is presented in Table~\ref{tab:cosmopara}. The power spectra of the density and velocity fluctuations are expressed as,
\ba
P_{\delta\delta}(k,z)&=& G^2_\delta(z){\cal G}(h;z){\cal A}(A^2_S,N_{\nu_0},N_\nu,\omega_{\nu};z_{\rm lss}){\cal P}_{\delta\delta}(k,z_{\rm lss})\nn\\
P_{\Theta\Theta}(k,z)&=& G^2_\Theta(z){\cal G}(h;z){\cal A}(A^2_S,N_{\nu_0},N_\nu,\omega_{\nu};z_{\rm lss}){\cal P}_{\Theta\Theta}(k,z_{\rm lss})
\nn
\ea
where the power spectra at last scattering surface with neutrino mass of emulator is given by,
\ba
P_{\delta\delta}(k,z_{\rm lss})&=& {\cal A}(A^2_S,N_{\nu_0},N_\nu,\omega_{\nu};z_{\rm lss}){\cal P}_{\delta\delta}(k,z_{\rm lss})\nn\\
P_{\Theta\Theta}(k,z_{\rm lss})&=& {\cal A}(A^2_S,N_{\nu_0},N_\nu,\omega_{\nu};z_{\rm lss}){\cal P}_{\Theta\Theta}(k,z_{\rm lss}),
\ea
where $\omega_{\nu}=\Omega_{\nu}h^2$. Here ${\cal G}$ denotes the isotropic volume factor and ${\cal A}$ the amplitude of the linear power spectrum at the last scattering surface, which differs from the emulator value when the neutrino mass differs. Therefore, the measured $f\sigma_8^*$ is corrected as 
\be
f\sigma_8=f\sigma_8^*\frac{{\cal G}(h;z){\cal A}(A^2_S,N_{\nu_0},N_\nu,\omega_{\nu};z_{\rm lss})}{{\cal G}(h^*;z){\cal A}(A^{2\,*}_S,N^*_{\nu_0},N^*_\nu,\omega^*_{\nu};z_{\rm lss})}
\ee
The superscript `$*$' denotes quantities estimated using the scale-independent parameters listed in the first column of Table~\ref{tab:cosmopara}, and thus ${\cal A}(A^{2\,*}_S,N^*_{\nu_0},N^*_\nu,\omega^*_{\nu};z_{\rm lss})=1$. The late-time \textit{scale-independent} factor ${\cal G}(h;z)/{\cal G}(h^*;z)$ is given by,
\be
{\cal G}(h;z)/{\cal G}(h^*;z)=\Delta D_V^{-1}
\ee
where $D_V=(D_A^2H^{-1})^{1/3}$ and $\Delta D_V$ denotes the deviation from the fiducial volume factor. This effect of the late-time \textit{scale-independent} parameter is applied at each step of the MCMC chain, and the correction factor applied to the measured $f\sigma_8^*$ is set by the combination of the volume factor ${\cal G}$ and the amplitude term ${\cal A}$. For $m_\nu=0.06\,{\rm eV}$ this factor is $[{\cal G}(h;z)/{\cal G}(h^*;z)]\times{\cal A}=1.005$, that is, a $0.5$ per cent correction.


\begin{figure}
\begin{center}
\includegraphics[width=\textwidth]{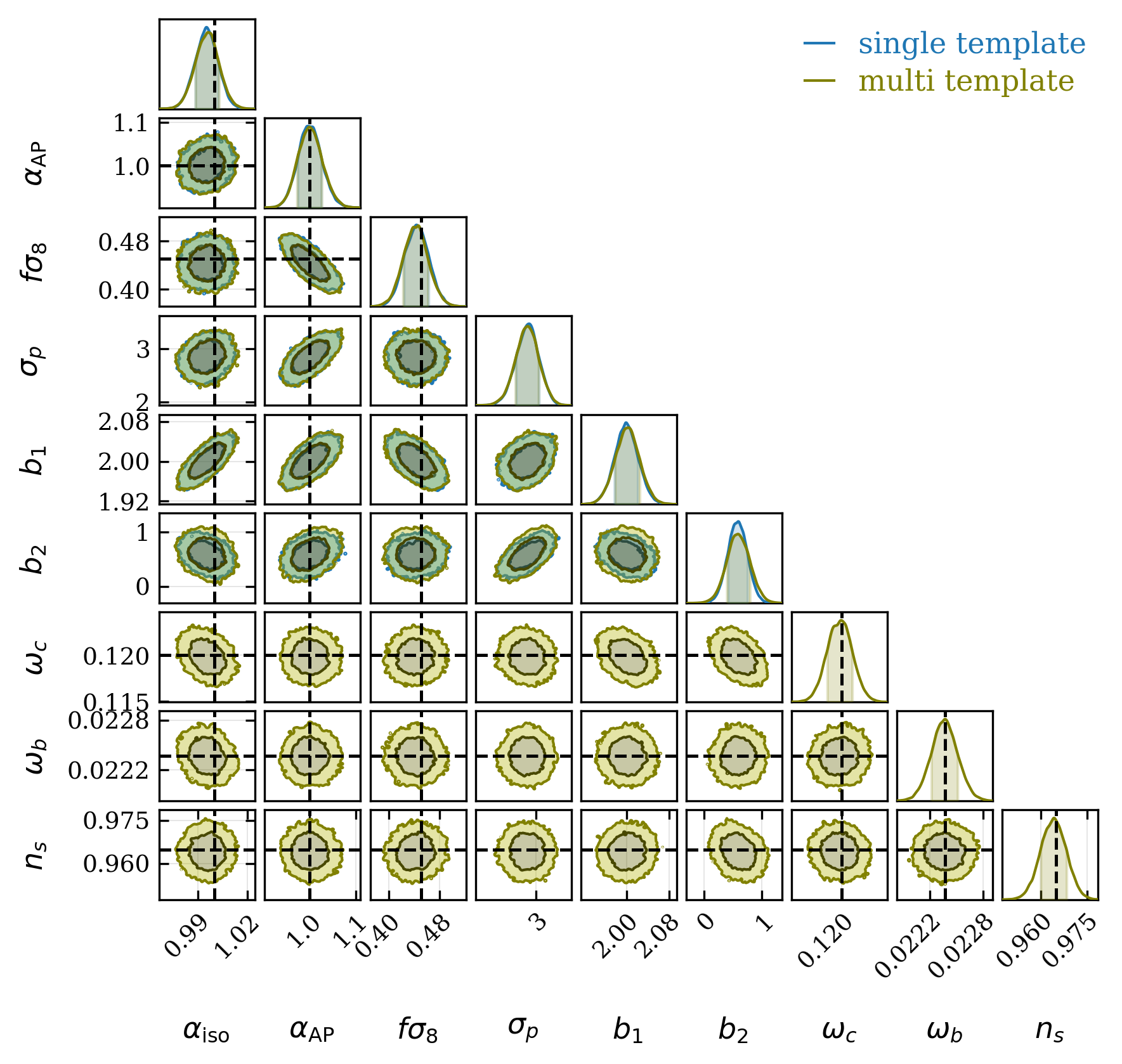}
\end{center}
\caption{{The corner plot recovered from the multi-template analysis of the mock catalogue (olive), compared with the single-template analysis of the same data (blue), showing the inner 68 and outer 95 per cent confidence regions. The error bars correspond to a single survey volume. The dilation parameters are given in the DESI convention, $\alpha_{\rm iso}=(\alpha_{||}\alpha_\perp^2)^{1/3}$ and $\alpha_{\rm AP}=\alpha_{||}/\alpha_\perp$, and are formed sample by sample from the chain. Dashed lines mark the input values of the simulation, and $\sigma_p$ is given in units of $\mpcoh$. The three shape parameters $(\omega_c,\omega_b,n_S)$ are sampled only in the multi-template fit, where they carry a Planck prior truncated at $4.5\sigma$ so that every chain element remains inside the range over which the emulator was trained.}} 
\label{fig:multi_corner_sim}
\end{figure}


\subsection{Verification test for Multi-template case}\label{subsec:multi_verify}

In this subsection, we introduce a multi-template analysis of the anisotropic power spectrum. The three primordial shape parameters $\omega_c$, $\omega_b$ and $n_S$ are added to the six model-independent parameters introduced in the previous subsection. While the influence of the \textit{scale-independent} parameters on the anisotropic power spectrum is easily estimated as a posterior correction by exploiting scaling relations, the impact of the \textit{scale-dependent} parameters becomes non-trivial beyond such simple scaling relations. 

We build the emulator so as to cover the impact of varying these \textit{scale-dependent} parameters on the power spectrum. The \textit{scale-dependent} parameter space is spanned up to $5\sigma$ of the Planck measurements for $\Lambda \rm{CDM}$, which are presented as $\omega_c=0.1202 \pm 5\times0.0014 $, $\omega_b=0.02236 \pm 5\times0.00015$ and $n_S=0.9649 \pm 5\times0.0044$. These parameter ranges are tight, but since the Planck prior is added to the $\chi^2$ estimation in the multi-template analysis, most of the chain elements remain within these boundaries. {In the analysis presented here the Planck prior is truncated at $4.5\sigma$, so that every chain element is guaranteed to lie inside the trained range rather than merely most of them.} We note that our emulators are constructed at the five  redshifts $z=0.3, 0.5, 0.7, 0.9, 1.1$. The power spectrum at any redshift is then estimated by interpolating among the predictions at these five redshifts. 

The measured parameters of the multi-template analysis are presented in the corner plot of Fig.~\ref{fig:multi_corner_sim}. These six parameters are shown after marginalizing over $\omega_c$, $\omega_b$ and $n_S$, which are allowed to vary freely. The three measured \textit{scale-dependent} parameters are not shown in this figure; they are found to follow essentially normal distributions centred on the Planck prior. This is because the simulations adopt cosmological parameters consistent with Planck. 

There is no significant deviation in the cosmic distance measures, but the error bars are slightly larger, at the sub-percent level. If \textit{scale-dependent} parameters different from the Planck prior were used for the simulations, we would observe deviations in $\omega_c$, $\omega_b$ and $n_S$, which would lead to results different from those of the single-template case. The observed consistency demonstrates that our analysis pipeline recovers the input cosmology within the precision of this particular validation test. 

In addition, the measurements of the growth factor $f\sigma_8$ are not significantly different, 
and the same holds for the FoG nuisance parameter $\sigma_p$. At least within the small variations of $\omega_c$, $\omega_b$ and $n_S$ allowed by the Planck prior, there is little difference even in the non-perturbative effect, which supports the validity of our simplified Gaussian FoG assumption under variations of the \textit{scale-dependent} parameters. We also note that the measured galaxy bias parameters $b_1$ and $b_2$ are again consistent with the values obtained in the single-template analysis.

{Quantitatively, with the error bars of a single survey volume the multi-template fit gives $\alpha_{\rm iso}=0.9956^{+0.0071}_{-0.0073}$, $\alpha_{\rm AP}=1.0013^{+0.0271}_{-0.0260}$ and $f\sigma_8=0.4424^{+0.0192}_{-0.0194}$, where the dilation parameters are quoted in the DESI convention, $\alpha_{\rm iso}=(\alpha_{||}\alpha_\perp^2)^{1/3}$ and $\alpha_{\rm AP}=\alpha_{||}/\alpha_\perp$, formed sample by sample from the chain. The single-template fit to the same data gives $\alpha_{\rm iso}=0.9952^{+0.0069}_{-0.0070}$, $\alpha_{\rm AP}=1.0008^{+0.0259}_{-0.0250}$ and $f\sigma_8=0.4439^{+0.0192}_{-0.0194}$. The posterior widths of the model-independent parameters differ by at most four per cent between the two fits, and the central values agree well within one standard deviation. The nuisance parameters behave in the same way, $b_1=2.0017\pm0.0230$, $b_2=0.594^{+0.203}_{-0.195}$ and $\sigma_p=2.85^{+0.20}_{-0.21}\,\mpcoh$, with only $b_2$ widening appreciably, by 18 per cent. The shape parameters are recovered at their input values, $\omega_c=0.1198\pm0.0013$, $\omega_b=0.02237\pm0.00015$ and $n_S=0.9641\pm0.0042$, each within $0.2\sigma$ of the prior centre and with a posterior width between 0.90 and 0.99 of the prior width. That last number is the honest summary of what the data add at this volume: the shape constraint is prior-dominated, which is precisely why opening it costs so little.}

{The cost of marginalising over the shape is not a fixed number, however; it depends on the strength of the data relative to the prior. Repeating the comparison against the covariance of the mean of the 25 realisations, which corresponds to 25 times the survey volume, the likelihood is no longer weak compared with the Planck prior and the shape directions begin to open. There the width of $\alpha_{\rm iso}$ grows by a factor of 1.83 when the shape is marginalised over, while $\alpha_{\rm AP}$ and $f\sigma_8$ remain unchanged to within one per cent. The asymmetry is the expected one, and it appears in the posterior as a correlation between $\alpha_{\rm iso}$ and $(\omega_c,\omega_b)$: $\alpha_{\rm iso}$ is a distance measured in units of the sound horizon, which those two parameters set, whereas the Alcock--Paczynski ratio and the velocity amplitude do not reference $r_d$. A survey substantially larger than DR1 will therefore pay for the shape marginalisation in the isotropic dilation first, and in the growth rate not at all.}

{The same 25-fold volume lets us ask where the model itself begins to fail, a question to which the survey-volume error bars are insensitive. Fitting to $k_{\rm max}=0.16\,h\,{\rm Mpc}^{-1}$ at that precision gives $\chi^2/{\rm dof}=145.4/47$, and the shape parameters no longer sit on their input values: $n_S$ is pulled $1.8\sigma$ below the prior centre. Lowering the scale cut to $k_{\rm max}=0.10\,h\,{\rm Mpc}^{-1}$ removes both symptoms at once, giving $\chi^2/{\rm dof}=23.5/23$ and every shape parameter within $0.3\sigma$ of its input. The pull on $n_S$ is therefore not a preference of the data, but the spectral tilt absorbing a modelling error confined to $k>0.10\,h\,{\rm Mpc}^{-1}$. At the volume of DR1 the same fit returns $n_S$ within $0.2\sigma$ of its input, so the effect is invisible in the measurements reported here; it indicates the precision at which this model will have to be improved, rather than a limitation of the present analysis.}

\section{Cosmological constraints with DESI DR1 data}\label{sec:data_analysis}

\begin{figure}
\begin{center}
\includegraphics[width=\textwidth]{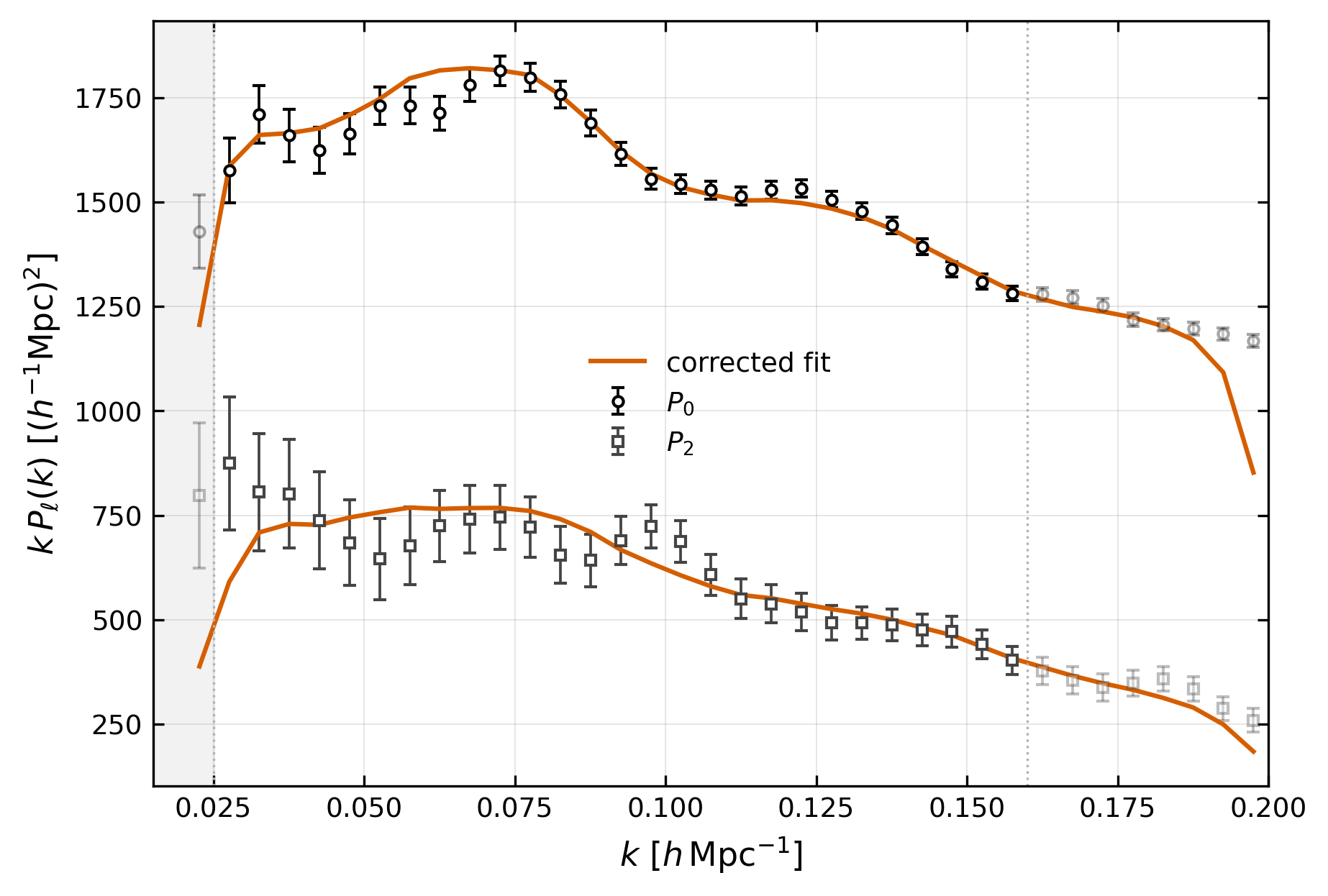}
\end{center}
\caption{The redshift-space power-spectrum multipoles measured from the DESI DR1 LRG2 sample. Black points show the monopole $P_0(k)$ and blue points the quadrupole $P_2(k)$, with error bars given by the square root of the diagonal elements of the covariance matrix. The solid and dashed curves are the corresponding best-fitting monopole and quadrupole. These are not the theoretical predictions themselves, but the predictions convolved with the DESI survey window matrix, so that the model and the measurements are compared on the same footing.} 
\label{fig:pkDESI}
\end{figure}

\subsection{Power-spectrum multipoles and covariance for the DESI DR1 LRG2 sample}

We analyse the anisotropic clustering of galaxies in the DESI DR1 LRG2 sample in the redshift interval $0.6 < z < 0.8$. The clustering information is extracted from measurements of the redshift-space power-spectrum multipoles introduced in \citep{2025JCAP...09..008A}. In this work, we use the monopole and quadrupole moments, $P_0(k)$ and $P_2(k)$, while the hexadecapole is not included in the analysis.

The power-spectrum measurements are derived from weighted galaxy overdensity fields constructed from the observed galaxy catalogue together with synthetic random catalogues that reproduce the angular and radial selection functions of the survey. The random catalogues correct for survey geometry, angular completeness, and observational selection effects, allowing an unbiased estimate of the large-scale clustering signal. We use the measurements provided by the DESI collaboration and adopt the standard DESI Fourier-space estimator and data products.

Because of the complex DESI survey footprint, the measured multipoles are convolved with the survey window function, which couples Fourier modes and mixes different multipole orders. To account for this effect, theoretical predictions are forward-modelled through the DESI window matrix before comparison with the data. This procedure ensures that the model and observations are defined on the same footing and consistently incorporates the impact of the survey geometry.

The covariance matrix is taken from the DESI DR1 Full-Shape clustering products and is primarily estimated from an ensemble of 1000 EZmock realisations, which reproduce both the clustering properties of the LRG sample and the DESI survey geometry. To account for residual uncertainties that are not captured by the mock-to-mock statistical variance, we additionally include the DESI systematic covariance matrix. This external covariance quantifies the impact of residual modelling and observational systematics, including uncertainties associated with the galaxy-halo connection, fibre-collision corrections, imaging and spectroscopic systematics, and the assumed fiducial cosmology. Following the DESI analysis methodology, the mock-derived covariance is corrected for the finite number of mock catalogues using the Hartlap \citep{Hartlap:2006kj} and Percival \citep{Percival:2013sga} corrections before being combined with the systematic covariance. {In the fits reported below the inverse covariance is the direct inverse of the sub-matrix of the fitted bins multiplied by $0.8948$, which is the correction carried by the inverse covariance supplied with these products; quoting it explicitly is what makes the $\chi^2$ values below reproducible. It scales $\chi^2$ by that factor and widens every error bar by $1/\sqrt{0.8948}=1.057$.}

\subsection{Measured values with a single-template case}
\label{subsec:DESI_single}
 
In this subsection, the model-independent parameter set $(\alpha_\perp,\,\alpha_{||},\,f\sigma_8,\,\sigma_p,\,b_1,\,b_2)$ is probed using the single-template method presented in Section~\ref{subsec:single_verify}. The monopole and quadrupole are fitted up to $k_{\rm max}=0.16\hompc$, {a choice the measurements are not sensitive to: at the volume of DR1 the recovered parameters move by within about one standard deviation as the cut is varied between $0.10$ and $0.20\hompc$, and Section~\ref{subsec:multi_verify} sets out what happens to the model itself at larger volumes.} The template is built on a fiducial cosmology with massless neutrinos, matching the cosmology on which our emulator was constructed, and the density fluctuation is normalized to {$\sigma_8=0.82$} at $z=0$. The primordial shape parameters are fixed to their Planck values, $\omega_c=0.1202$, $\omega_b=0.02236$ and $n_S=0.9649$. {The verification test against simulations in Section~\ref{subsec:single_verify} recovers the input cosmology to within $0.7\sigma$ at one survey volume and $1.7\sigma$ at five, the two volumes at which the DESI full-shape analysis validates its own pipeline; the half-per-cent offset that becomes visible at larger volumes is a third of the statistical uncertainty of the measurement reported here.} {The fit uses the range $0.025<k<0.16\hompc$ for both multipoles, $54$ data points in total. The lowest measured bin, at $k=0.0225\hompc$, is excluded. It is the bin in which the window matrix suppresses the power most strongly, and the measurement there does not lie where the convolved model places it; since $0.0225\hompc$ is not a scale at which non-linear modelling is in question, we interpret this as a systematic of the measurement rather than a failure of the model, and remove it from the fit. The resulting best fit has $\chi^2=38.0$ for $48$ degrees of freedom. The measured and the best fit multipoles are presented in Fig~\ref{fig:pkDESI}.}

\begin{figure}
\begin{center}
\includegraphics[width=0.710\textwidth]{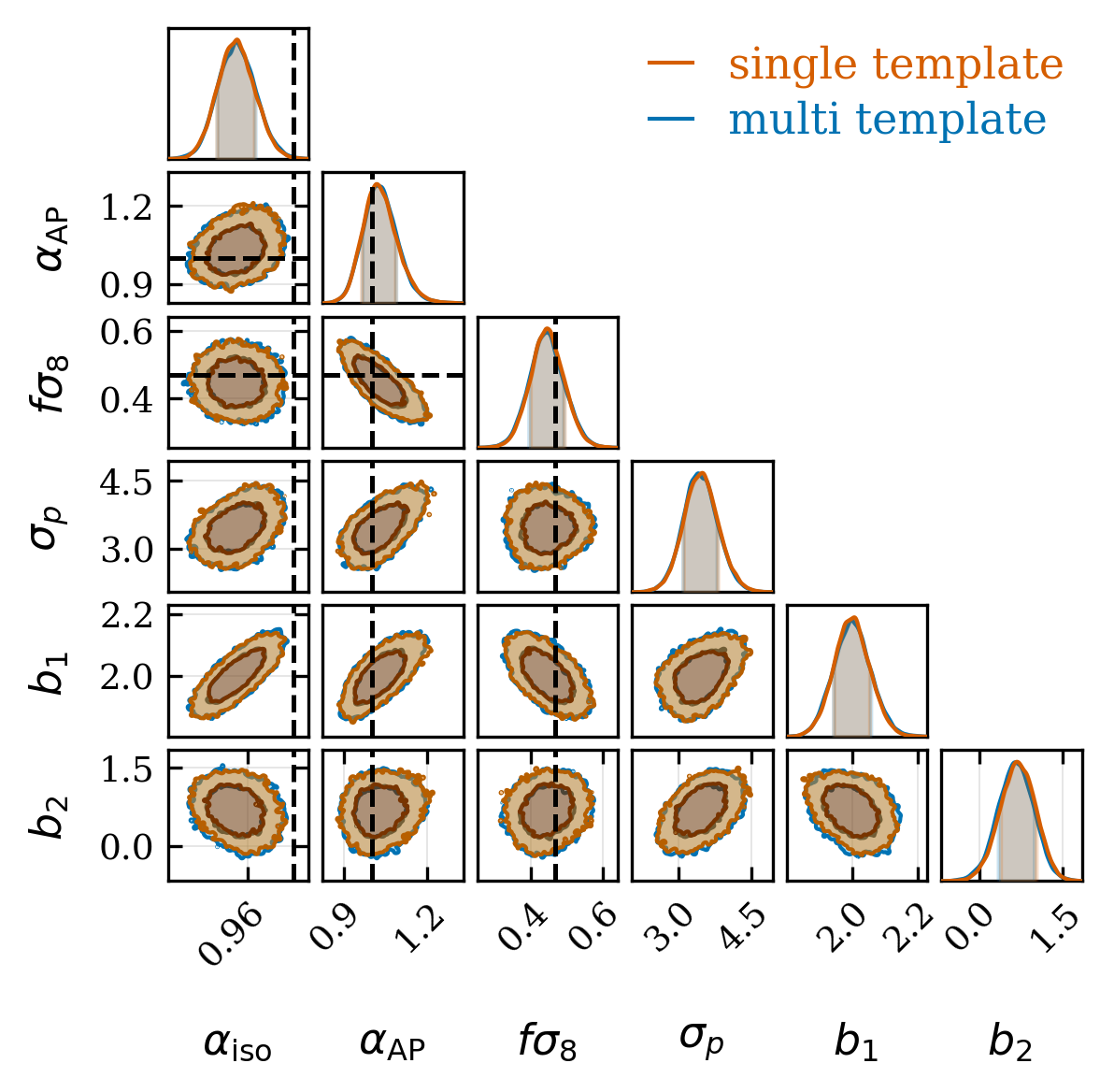}
\end{center}
\caption{{The corner plot for the model-independent parameters $(\alpha_{\rm iso},\alpha_{\rm AP},f\sigma_8,\sigma_p,b_1,b_2)$ recovered from the DESI DR1 LRG2 sample, showing the inner 68 and outer 95 per cent confidence regions. Orange shows the single-template analysis, in which the shape parameters are held at the Planck values, and blue the multi-template analysis, in which $\omega_{\rm c}$, $\omega_{\rm b}$ and $n_S$ vary under the Planck prior and are marginalised over here. The two sets of contours are almost indistinguishable, which is the central result of this section. Dashed lines mark the fiducial values, and $\sigma_p$ is given in units of $\mpcoh$.}} 
\label{fig:multi_template_DESI}
\end{figure}

{The cosmic distance measures are constrained to $\alpha_\perp=0.940^{+0.021}_{-0.021}$ and $\alpha_{||}=0.967^{+0.049}_{-0.042}$, or equivalently $\alpha_{\rm iso}=0.950^{+0.017}_{-0.016}$ and $\alpha_{\rm AP}=1.028^{+0.066}_{-0.057}$ in the DESI convention, and their two-dimensional posterior is shown in the corner plot of Fig.~\ref{fig:multi_template_DESI}. The isotropic dilation is measured at the $1.7$ per cent level, while the Alcock-Paczynski ratio is constrained nearly four times less tightly. Both distance measures fall below the fiducial cosmology, $\alpha_{\rm iso}$ by about $3\sigma$. This is not peculiar to our analysis. The DESI full-shape analysis of the same sample \citep{2025JCAP...09..008A}, converted to our variables as described at the end of Section~\ref{subsec:DESI_multi}, gives $\alpha^{\rm DESI}_\perp=0.945\pm0.027$ and $\alpha^{\rm DESI}_{||}=0.967\pm0.053$, which agree with our values to well within $1\sigma$ with uncertainties about $30$ per cent larger. The two analyses share the measurement and the covariance but differ in the model, so this agreement speaks to the robustness of the distance measurement against the choice of model rather than providing an independent confirmation of it. Nor is the feature new: the DESI analysis itself reports that both its full-shape and its baryon-acoustic-oscillation measurements of the LRG2 sample return an isotropic dilation two to three standard deviations below the fiducial value, and figure 13 of that reference and the discussion around it are devoted to the point. {The agreement with the DESI full-shape results, despite the different theoretical modelling approaches, suggests that the preference for a low $\alpha_{\rm iso}$  in the LRG2 sample is not an artefact of any single modelling choice.} }

The growth factor is constrained to {$f\sigma_8=0.445^{+0.048}_{-0.046}$}, assuming the minimal neutrino mass of $m_\nu = 0.06\,{\rm eV}$, which is consistent with the fiducial value of $f\sigma_8=0.467$ at the effective redshift $z=0.7$ of the LRG2 sample. The measured dispersion of {$11\%$} is, however, considerably larger than the theoretical expectation of {$5\%$} obtained from a Fisher calculation {\citep{2015JCAP...08..007S}}, which assumes a perfectly Gaussian posterior. Two factors plausibly contribute to this difference. First, the Fisher forecast represents an idealized limit, while the posterior obtained here is visibly asymmetric, with the lower bound extending further than the upper one. Second, the forecast is evaluated for the completed survey, whereas the present analysis is based on DR1, which covers only part of the planned survey volume. We discuss the FoG dispersion parameter $\sigma_p$ together with the growth factor here, since it is the parameter most strongly degenerate with $f\sigma_8$ and both are governed by the peculiar velocity field. It is measured to be {$\sigma_p=3.46^{+0.36}_{-0.35}\mpcoh$, slightly larger than the linear velocity dispersion of the template, $3.30\mpcoh$, obtained by integrating its linear velocity-divergence spectrum}, though the difference is well within the uncertainty. Such a mild excess is expected, since the scales targeted in our analysis lie in the mildly quasi-linear regime, where non-linear contributions to the velocity field are already present.

Given the target halo mass of the DESI galaxy sample, the leading-order galaxy bias is expected to be around $b_1=2$. Although $b_1$ is a nuisance parameter that must be marginalized over when probing the key cosmological parameters, its measured value provides a useful check on the analysis: a large departure from this expectation would indicate a problem with the measurement. We obtain {$b_1=2.000^{+0.055}_{-0.053}$} from the data, {while the same analysis applied to the AbacusSummit LRG2 mock catalogues described in Section~\ref{subsec:single_verify} yields $b_1=2.00$}. The agreement between the two supports the validity of our measurement. The second-order galaxy bias is constrained to {$b_2=0.69^{+0.32}_{-0.32}$}. The positive value enhances the power spectrum on quasi-linear scales rather than suppressing it, although the deviation from zero is only at the $2\sigma$ level.

Finally, we note that our analysis has no sensitivity to the neutrino mass itself. Neutrino oscillation experiments imply a minimum total mass of $m_\nu =0.06$eV, and the $f\sigma_8$ reported above is estimated assuming a single massive neutrino with this minimal value. For such a small mass, the effect of the neutrino is an overall damping of the amplitude, which is completely degenerate with the growth parameter. If neutrinos were massless, the measured $f\sigma_8$ would be $0.5\%$ smaller and the normalization would become $\sigma_8=0.808$; apart from $b_1$, the remaining measurements would be largely unaffected. We therefore apply this $0.5\%$ shift to the central value of $f\sigma_8$ in our final measurements, leaving the uncertainties unchanged. These single-template results are compared with those obtained from the multi-template analysis in Section~\ref{subsec:DESI_multi}.

\subsection{Measured values with a Multi-template case}\label{subsec:DESI_multi}

In this subsection, the model-independent parameter set is extended by the three shape parameters, $(\alpha_\perp,\,\alpha_{||}, \,f\sigma_8,\,\sigma_p,\,b_1,\,b_2,\,\omega_{\rm c},\,\omega_{\rm b},\,n_S)$, and is measured using the multi-template method presented in Section~\ref{subsec:multi_verify}. As in the single-template analysis, only the monopole and quadrupole are used, the same threshold of $k_{\rm max}=0.16\hompc$ is adopted, and the normalization is set to {$\sigma_8=0.82$} at $z=0$ for the massless-neutrino cosmology. 
{Instead of being fixed, the shape parameters $\omega_c$, $\omega_b$ and $n_S$ are varied under the Planck constraints of $\omega_c=0.1202\pm 0.0014$, $\omega_b=0.02236 \pm 0.00015$ and $n_S=0.9649 \pm 0.0044$, truncated at $4.5\sigma$ so that every sample remains inside the training box of the emulator.}

{The shape posteriors we recover are $\omega_c=0.1205\pm 0.0014$, $\omega_b=0.02236\pm 0.00015$ and $n_S=0.9655\pm 0.0045$. These reproduce the prior: their widths are $1.00$, $1.01$ and $1.01$ times the prior widths, and their centres are displaced from the prior centres by $+0.21\sigma$, $-0.02\sigma$ and $+0.15\sigma$. The monopole and quadrupole below $k_{\rm max}=0.16\hompc$ therefore add no information on the shape beyond that already supplied by Planck. The correct reading of this analysis is accordingly not that the shape has been measured from DESI, but that the measurement of distance and growth has been marginalised over the entire range of templates the Planck prior allows, and is unaffected by it. Asking whether DESI and Planck prefer the same shape would require a looser prior, and in our framework also an emulator trained over a correspondingly wider box; we leave that to future work.} 

{The cosmic distance measures are constrained to $\alpha_{\rm iso}=0.950^{+0.017}_{-0.016}$ and $\alpha_{\rm AP}=1.030^{+0.064}_{-0.058}$, against $\alpha_{\rm iso}=0.950^{+0.017}_{-0.016}$ and $\alpha_{\rm AP}=1.028^{+0.066}_{-0.057}$ in the single-template case. The central values shift by less than a twentieth of the statistical uncertainty and the errors change by a few per cent at most, so the two sets of contours in Fig.~\ref{fig:multi_template_DESI} lie on top of one another. The correlations between the shape parameters and the distance measures are weak, the strongest being $-0.14$ between $\alpha_{\rm iso}$ and $\omega_c$, which is why opening the three shape directions neither displaces the central values nor inflates the errors. }

{The growth factor is constrained to $f\sigma_8=0.444^{+0.050}_{-0.047}$, consistent with the single-template value of $0.445^{+0.048}_{-0.046}$ and, like the distance measures, shifted by far less than its own uncertainty. The non-perturbative velocity dispersion is measured to be $\sigma_p=3.45^{+0.36}_{-0.34}\mpcoh$, again consistent with the single-template case.}

The nuisance parameters $b_1$ and $b_2$ are fully marginalized over when measuring the other cosmological parameters. Their best-fit values and uncertainties are consistent with those of the single-template fit, which supports that the cosmological measurements are reliable in multi-template case. {They are constrained to be $b_1=2.001^{+0.057}_{-0.056}$ and $b_2=0.66^{+0.32}_{-0.32}$. The best fit has $\chi^2=38.0$ for $45$ degrees of freedom, the same minimum as the single-template fit reaches with three fewer parameters. All of the measurements of this section are collected in Table~\ref{tab:DESIresults}. To summarise, the shape parameters are returned at their prior and the model-independent parameters are unchanged, so the distance and growth measurements of Section~\ref{subsec:DESI_single} hold when the template is allowed to vary over the full range Planck permits.}

{\begin{table}
\renewcommand{\arraystretch}{1.5}
\setlength{\tabcolsep}{0.10cm}
\begin{center}
\footnotesize
\resizebox{\columnwidth}{!}{%
\begin{tabular}{|l|c|c|c||c|c|c||c|c|c||c|}
\hline
\hline
 & $\alpha_{\rm iso}$ & $\alpha_{\rm AP}$ & $f\sigma_8$ & $\sigma_p\,[\mpcoh]$ & $b_1$ & $b_2$ & $\omega_{\rm c}$ & $\omega_{\rm b}$ & $n_S$ & $\chi^2/{\rm dof}$ \\
\hline
\hline
Planck prior & -- & -- & -- & -- & -- & -- & $0.1202$ & $0.02236$ & $0.9649$ & \\
 &  &  &  &  &  &  & $\pm0.0014$ & $\pm0.00015$ & $\pm0.0044$ & \\
\hline
Single template & $0.950^{+0.017}_{-0.016}$ & $1.028^{+0.066}_{-0.057}$ & $0.445^{+0.048}_{-0.046}$ & $3.46^{+0.36}_{-0.35}$ & $2.000^{+0.055}_{-0.053}$ & $0.69^{+0.32}_{-0.32}$ & $0.1202$ & $0.02236$ & $0.9649$ & $38.0/48$ \\
 &  &  &  &  &  &  & (fixed) & (fixed) & (fixed) & $=0.79$ \\
\hline
Multi template & $0.950^{+0.017}_{-0.016}$ & $1.030^{+0.064}_{-0.058}$ & $0.444^{+0.050}_{-0.047}$ & $3.45^{+0.36}_{-0.34}$ & $2.001^{+0.057}_{-0.056}$ & $0.66^{+0.32}_{-0.32}$ & $0.1205$ & $0.02236$ & $0.9655$ & $38.0/45$ \\
 &  &  &  &  &  &  & $\pm0.0014$ & $\pm0.00015$ & $\pm0.0045$ & $=0.84$ \\
\hline
\hline
\end{tabular}}
\end{center}
\caption{Measurements from the DESI DR1 LRG2 sample, for the single-template analysis of Section~\ref{subsec:DESI_single} and the multi-template analysis of Section~\ref{subsec:DESI_multi}. Both fit the monopole and quadrupole over $0.025<k<0.16\hompc$, $54$ data points, with the same measurement, window matrix and covariance; they differ only in whether the three shape parameters are held at the Planck values or varied under the Planck prior. Errors are the $68$ per cent credible intervals of the marginalised posteriors, and $\alpha_{\rm iso}$ and $\alpha_{\rm AP}$ are formed sample by sample rather than from the medians. The quoted $f\sigma_8$ is the physical velocity amplitude: the chain parameter is $f\sigma_8\sqrt{A_s/A_s^*}$ and is divided by that factor and multiplied by the $0.5$ per cent neutrino correction of Section~\ref{subsec:simulations} before being quoted. The inverse covariance carries the finite-realisation correction of Section~\ref{sec:data_analysis}. The first row repeats the Planck prior, which makes it plain that the shape posteriors of the multi-template fit reproduce it.}
\label{tab:DESIresults}
\end{table}
}

Because \cite{2025JCAP...09..008A} constrained distances and growth alongside two additional shape parameters ($m,n$), it is instructive to compare their results with our own. While we physically model the broadband shape using cosmological parameters ($\omega_c$, $\omega_b$, and $n_S$), their analysis employs a phenomenological parameterization. To enable a direct comparison, we focus on the geometric and growth parameters ($\alpha_{\parallel}, \alpha_{\perp}, f\sigma_8$) marginalizing over the respective shape parameters in both analyses. Using the constraints from Table 9 of \cite{2025JCAP...09..008A} and the covariance framework from Eq. 1.6 of \cite{2026JCAP...02E..02A}, we convert their reported $\alpha_{\rm iso}$ and $\alpha_{\rm AP}$ values into $\alpha_{\parallel}, \alpha_{\perp}$ via Eqs. 4.7 and 4.8. This transformation yields derived DESI constraints of $\alpha^{\rm DESI}_\perp=0.945\pm0.027$, $\alpha^{\rm DESI}_{||}=0.967\pm0.053$ and $f\sigma^{\rm DESI}_8=0.484\pm0.055$, which agree with our own to well within one standard deviation. Where the original posterior distributions for the A-P parameters were asymmetric, we have conservatively adopted the larger uncertainty bounds for this comparison.

\section{Conclusions}
\label{sec:Conclucion}

{Measuring distances and the growth rate without reference to a cosmological model is an old idea; what has changed is the precision now demanded of it. At the level DESI reaches, a description of redshift-space clustering accurate to a per cent on quasi-linear scales cannot be written down in closed form, and some form of simulation-calibrated emulation becomes unavoidable. This is where the difficulty lies. An emulator is trained on a grid of cosmologies, and a cosmology is specified by its parameters; no simulation can be run in $\alpha_\perp$, $\alpha_{||}$ and $f\sigma_8$, because these are not inputs to a simulation but properties of the comparison between one cosmology and another. An emulator built directly in the model-independent variables therefore cannot be constructed, and this is the obstacle that has to be removed before a model-independent analysis can be carried out at emulator precision.

The hybrid construction presented in this paper is our answer to it. The model-independent parameters enter analytically, through the perturbation-theory expressions of Section~\ref{sec:HybridModel}, where their meaning is exact by construction and no training is involved. The emulator is asked only for what perturbation theory cannot supply, the residual between the resummed propagators and the $N$-body result, and it is trained over the shape parameters $\omega_{\rm c}$, $\omega_{\rm b}$ and $n_S$, which are inputs to a simulation. It is this division of labour that makes a precise measurement of the model-independent parameters possible while the shape of the power spectrum is varied at the same time.

It is worth being explicit about how this differs from the two families of method currently in use. The first keeps the compression that classical analyses apply and extends it, adding effective parameters that carry the shape information. In ShapeFit \citep{Brieden:2021edu,Brieden:2021cfg} there are two of them: a slope parameter $m$, defined as the response of the broadband to a change of cosmology evaluated around a fiducial point, and a parameter $n$ that carries the primordial tilt, the part played by $n_S$ in our own set. The construction is phenomenological by design, and deliberately so: $m$ is not a cosmological parameter but a compression of the effect of several, and its adequacy is something to be established rather than assumed. The comparisons carried out for DESI find the compressed and direct routes to agree at the $0.5\sigma$ level within $\Lambda$CDM, with the differences growing for extended models. Our shape parameters are instead $\omega_{\rm c}$, $\omega_{\rm b}$ and $n_S$ themselves, varied without compression, and the accuracy of the model at each point in that space is supplied by an emulator trained at that point rather than by a response calibrated elsewhere. The two approaches answer different questions, and the direct one becomes the relevant question as soon as the prior on the shape is loosened, since a response linearised about a fiducial cosmology is of limited use far from it.

The second family fits the full shape with an effective field theory of large-scale structure \citep{Ivanov:2019pdj,DAmico:2019fhj}. These analyses do vary the cosmology and they are accurate; our purpose is not to offer a replacement for them. The point is rather that the effective field theory is one modelling choice among several, and that its accuracy on quasi-linear scales is carried by counterterms and stochastic parameters that are free and are fitted to the same data. A measurement whose accuracy comes instead from perturbation theory with $N$-body-calibrated residuals, and which is written from the outset in model-independent variables, reaches the same data from a different direction. Where two routes that differ this much agree, the agreement constrains the modelling error in a way that neither can on its own; where they disagree, the disagreement locates the assumption responsible.

Two such comparisons are available here. The first is with the compressed measurement of the same sample reported in \citep{2025JCAP...09..008A}, which agrees with ours to well within $1\sigma$, as set out at the end of Section~\ref{subsec:DESI_multi}. The second is with the effective-field-theory fit itself. That analysis reports $\Lambda$CDM parameters rather than distances, but its cosmology can be brought into our variables by asking what it predicts at the effective redshift of LRG2: $\Omega_m=0.296$ and $H_0=68.63\,{\rm km\,s^{-1}\,Mpc^{-1}}$ with $\sigma_8=0.841$ give $\alpha_{\rm iso}=0.991$, $\alpha_{\rm AP}=1.008$ and $f\sigma_8=0.476$ at $z=0.7$. Against our multi-template measurement this is $\chi^2=7.2$ for three degrees of freedom, and the discrepancy sits almost entirely in $\alpha_{\rm iso}$, which differs by $2.5\sigma$; the Alcock-Paczynski ratio and the growth rate agree to within $0.3\sigma$ and $0.7\sigma$. We have not propagated the uncertainties of their constraint, which would move the comparison towards agreement. The difference should not be read as one between the two models. Their constraint is obtained from six tracers jointly and ours from one, and the compressed measurement of that same tracer sits where ours does rather than where the joint fit does, so what the comparison shows is this sample against the ensemble. The model-independent parameterisation is what makes that visible at all: a difference expressed in $\Omega_m$ and $H_0$ cannot say which of the two distances is responsible, and here it is the isotropic one alone.

Section~\ref{sec:data_analysis} shows what the method returns on real data. Letting the three shape parameters vary over the range the Planck prior permits leaves the model-independent measurements where the single-template fit put them: the central values move by less than a twentieth of their own uncertainties, the errors change by a few per cent at most, and the best fit reaches the same $\chi^2$ with three more free parameters. The shape posteriors come back at the prior. This is a null result and is worth stating plainly as one. Below $k=0.16\hompc$ the DESI DR1 LRG2 monopole and quadrupole carry no information about the shape beyond what Planck already supplies, and the distance and growth measurements of Section~\ref{subsec:DESI_single} are insensitive to the template freedom that remains within it.

Because they are model-independent, these measurements can be confronted with a cosmological model after the fit rather than during it. Within flat $\Lambda$CDM with $\omega_{\rm c}$ and $\omega_{\rm b}$ held at the Planck values, the only freedom left is $h$: $\Omega_m=(\omega_{\rm c}+\omega_{\rm b})/h^2$ then fixes $D_A(z)$, $H(z)$ and $f\sigma_8(z)$ together. Fitting that single parameter to the three measurements $(\alpha_{\rm iso},\alpha_{\rm AP},f\sigma_8)$, with the $3\times3$ covariance taken from the chain so that the strong $\alpha_{\rm AP}$--$f\sigma_8$ anticorrelation is carried properly, gives $h=0.724\pm0.017$ and $\Omega_m=0.272$ with $\chi^2=0.7$ for two degrees of freedom from the multi-template chain, and $h=0.724\pm0.016$ with $\chi^2=0.7$ from the single-template chain. $\Lambda$CDM thus describes the three measurements well. The test is informative in the other direction: the fiducial $h=0.6727$ is disfavoured at $\chi^2=11$, so what this sample prefers is the same model with a higher expansion rate.

The value is of some interest in itself. Read as an expansion rate it is $H_0=72.4\pm1.7\,{\rm km\,s^{-1}\,Mpc^{-1}}$, which sits within $1\sigma$ of the local distance-ladder determination $H_0=73.0\pm1.0\,{\rm km\,s^{-1}\,Mpc^{-1}}$ \citep{Riess:2021jrx} and about $3.3\sigma$ above the value that defines our fiducial cosmology. We are careful about what this is. It is not an independent measurement of $H_0$: the absolute scale is set by the Planck shape parameters, which also fix the sound horizon, so this is a calibrated standard-ruler inference and not a rung of the distance ladder. Nor is it a cosmological constraint, coming as it does from one tracer in one redshift interval; the global constraint is formed from the combination of redshift bins and tracers, and the degeneracy between $h$ and $\omega_m$ is broken there and not here. What can be said is that the distances this sample prefers are those of a $\Lambda$CDM universe expanding faster than the Planck one, that the DESI full-shape analysis of the same sample sits at the same distances, and that the coincidence with the local value is close enough to be worth recording. The reason we are able to record it at all is that the measurement was made without reference to a cosmological model, so the comparison could be made afterwards and at no cost to the fit.}

{One number deserves a comment of its own. Fits of physically motivated models to redshift-space clustering have for a long time returned a velocity amplitude below the input value of the simulations against which they were tested. With the emulator the residual deficit at $k_{\rm max}=0.16\,h\,{\rm Mpc}^{-1}$ amounts to $f\sigma_8=0.444\pm0.019$ against an input value of $f\sigma_8=0.4506$ for the mock at $z=0.8$, that is, agreement to a third of a standard deviation at the precision of DR1. We regard reaching that level with a model built from perturbation theory and calibrated $N$-body residuals, rather than from free effective parameters, as the principal gain of the approach presented here. Models that place the velocity amplitude on the input value more exactly are in general effective descriptions whose additional freedom is itself fitted to the same simulations, and we have deliberately not used them: a parameter able to absorb the discrepancy cannot distinguish a correct model from a tuned one.}

{We are, however, explicit about what the scale-cut test of Section~\ref{subsec:multi_verify} does not settle. Restricting the fit to $k<0.10\,h\,{\rm Mpc}^{-1}$ returns the shape parameters to their input values and brings the reduced $\chi^2$ to unity even at 25 times the survey volume, but it does not remove the velocity-amplitude deficit; measured against the covariance of the mean of the 25 realisations the deficit is in fact larger at the lower cut. A fit restricted to quasi-linear scales is therefore not simply a cleaner version of the same measurement, and the possibility that agreement in the linear growth rate is bought at the price of the non-linear description is one to be tested rather than assumed. That test belongs to the next revision of the emulator.}

{We can be more specific about where the model runs out. The redshift-space mapping of Section~\ref{sec:HybridModel} carries its higher-order corrections as a polynomial in $k\mu$, truncated at the order to which they have been computed, and the data want more of that term than the truncation provides. Scaling the correction by a constant and refitting at twenty-five times the survey volume, the velocity amplitude reaches the input value of the simulation at a scaling of about $1.5$, and the $\chi^2$ of a fit with the cosmology pinned at the simulation values is minimised at the same place; the dilation along the line of sight crosses unity at about $2$. That two quantities not tied to one another are optimised together is not how a nuisance term absorbing a mismatch behaves. Beyond that scaling the six-parameter $\chi^2$ keeps falling while the pinned $\chi^2$ rises steeply, so the further improvement is absorption rather than physics, and the apparent preference of the free fit for a much larger correction should not be read as evidence for one. Even at the optimum the pinned fit is poor, $\chi^2=146$ for $53$ degrees of freedom, so the amplitude of this term is not the only thing the model is missing at that volume. Whether the deficit is supplied by the next order of the expansion or requires a resummation of it is not settled by this test, and answering that question is what the next revision of the model has to do.}

{Is the method then no longer needed, having returned a null result? We would argue the opposite. The null result is a statement about the Planck prior and not about the shape: it says that the template freedom Planck allows is too small to matter, which is itself something that had to be demonstrated rather than assumed, and which we could not have demonstrated without the construction described here. Put as a number rather than as a null, the choice of template contributes less than a twentieth of the statistical uncertainty to the error budget of the DESI DR1 LRG2 distance and growth measurements, as long as the shape is held within the Planck prior. Analyses that fix the template have until now had to assume something of this kind; here it is measured. The question that follows is the one the present analysis cannot yet answer, namely which shape of the power spectrum the DESI observations themselves prefer, measured simultaneously with the distances and the growth rate and without borrowing the answer from the cosmic microwave background. That measurement needs exactly this machinery, extended to a training box wide enough to contain a loosened prior, and it is the subject of the paper we are now preparing. The present work is the step that makes it possible: it sets out the model, it verifies against simulations that the model-independent parameters are recovered without bias when the template is allowed to move, and it quantifies how much of the current answer the prior is supplying.}

\section{Acknowledgements}
MJ is grateful to Yu Yu and Zhao Chen for useful discussions and comments on simulations. Y.Z. acknowledges the support from the National SKA Program of China (2025SKA0150104) and the National Natural Science Foundation o fChina (NFSC) through grant 12203107. 
MIL has been supported by the Atracción de Talento Contract no. 2023-5A/TIC-
28943 granted by the Comunidad de Madrid in Spain and CNS2024-154242
from the Spanish national research council (AEI). This work was supported by the high performance computing cluster Seondeok at the Korea Astronomy and Space Science Institute. This work was supported by the major research project No. 2026183200 of the Korea Astronomy and Space Science Institute, funded by the Ministry of Science and ICT.



\appendix

\section{High precision modelling of the density and velocity matter power spectrum}
\label{appendix:hybrid}

This section details our procedure for computing the two-loop theoretical templates, $\bar{P}^{PT}_{XY}(k, z)$, introduced in Eq.~\ref{eq:residual}. We first formulate the template evaluation within a fiducial cosmology—denoted consistently by barred variables—before establishing the scaling relations required to map these predictions onto target cosmologies.Following the RegPT framework developed by \cite{2012PhRvD..86j3528T}, the power spectrum is expanded using multi-point propagators. Up to the two-loop order, this expansion is given by:

\ba
\label{therietical_PXY}
&&\bar P^{PT}_{XY}(k,z) = \bar \Gamma_X^{(1)}(k,z)\bar \Gamma_Y^{(1)}(k,z)\bar P_{\rm lin}(k)
\nn
\\
&&\quad
+2\int\frac{d^3\vec q}{(2\pi)^3}\bar \Gamma_X^{(2)}(\vec q, \vec k-\vec q,z) \bar \Gamma_Y^{(2)}(\vec q, \vec k-\vec q,z)
\nn
\\
&&\quad\quad\quad
\times \bar  P_{\rm lin}(q) \bar  P_{\rm lin}(|\vec k-\vec q|) 
\nn
\\
&&\quad
+6\int\frac{d^3\vec pd^3\vec q}{(2\pi)^6}
\nn
\\
&&\quad\quad\quad
\times \bar \Gamma_X^{(3)}(\vec p,\vec q, \vec k-\vec p-\vec q,z)\bar \Gamma_Y^{(3)}(\vec p,\vec q, \vec k-\vec p-\vec q,z) 
\nn
\\
&&\quad\quad\quad
\times\bar  P_{\rm lin}(p) \bar  P_{\rm lin}(q) \bar  P_{\rm lin}(|\vec k-\vec p-\vec q|),
\label{eq:pk_RegPT}
\ea
where $\bar \Gamma_X^{(p)}$ represents the regularised $(p+1)$-point propagator. Evaluating Eq.~\ref{eq:pk_RegPT} therefore requires estimating $\bar \Gamma_X^{(p)}$ for $p=1,2,3$.   At the two-loop order, the two-point propagator ($p=1$) takes the form:
\ba\label{eq:gammax1}\bar \Gamma_X^{(1)}(k,z) = {\rm exp}\left(-\bar G_\delta^2 \bar\gamma \right)\sum_n \bar G_X \bar G_\delta^{n-1} \bar{\cal C}^{(1)}(\bar \gamma),
\ea
where $\bar\gamma=k^2\bar \sigma^2_{\rm d}/2$. The term $\bar \sigma_{\rm d}^2$ represents the displacement field dispersion, computed from the initial linear power spectrum as $\bar \sigma_{\rm d}^2=\int_0^{k/2} (dq/6\pi^2)\bar P_{\rm lin}(q)$. The coefficients $\bar{\cal C}^{(1)}_n$ are derived from standard perturbation theory (PT). They strictly vanish for even values of $n$. For odd values, accounting for theoretical uncertainties, they are explicitly given by:
\ba
\bar{\cal C}^{(1)}_1(\bar \gamma) &=&1, \nn\\
\bar{\cal C}^{(1)}_3(\bar \gamma) &=& \bar \gamma + \bar \Gamma_{X,{\rm 1-loop}}^{(1)}(k),\nn\\
\bar{\cal C}^{(1)}_5(\bar \gamma) &=& \bar \gamma^2/2 + \bar \gamma\bar \Gamma_{X,{\rm 1-loop}}^{(1)}(k) + \bar \Gamma_{X,{\rm 2-loop}}^{(1)}(k)+\bar{\cal O}^{(1)}_{X,5},\nn\\
\bar{\cal C}^{(1)}_n(\bar \gamma) &=& \bar{\cal O}^{(1)}_{X,n}\,.
\ea
Here, $\bar \Gamma_{X,{\rm n-loop}}^p$ denotes the standard $n$-loop $(p+1)$-point PT propagator, whose exact forms are detailed in \cite{2012PhRvD..86j3528T,PhysRevD.89.023502}, distinguishing them from their regularised counterparts. The residual terms, $\bar{\cal O}^{(1)}_{X,n}$, capture PT systematics and uncertainties, which we calibrate against N-body simulations. We assume these theoretical uncertainties stem from both the two-loop and three-loop orders, a consequence partially driven by the UV-sensitivity of single-stream PT calculations.The corresponding expressions for the three-point propagator ($p=2$) are:
\ba
\bar \Gamma_X^{(2)}(k,z) &=& {\rm exp}\left(-\bar G_\delta^2\,\bar\gamma \right)\sum_n \bar G_X\,\bar G_\delta^{n-1} \bar{\cal C}^{(2)}_n\,\,;\nn \\
\bar{\cal C}^{(2)}_2(\bar \gamma) &=& \bar F_X^{(2)}(\vec q, \vec k-\vec q),\nn\\
\bar{\cal C}^{(2)}_4(\bar \gamma) &=& \bar\gamma \bar F_X^{(2)}(\vec q, \vec k-\vec q)+\bar\Gamma_{X,{\rm 1-loop}}^{(2)}(\vec q, \vec k-\vec q)+\bar{\cal O}^{(2)}_{X,4}\,, \nn\\ 
\bar{\cal C}^{(2)}_n(\bar \gamma) &=& \bar{\cal O}^{(2)}_{X,n}\,,
\ea
with $\bar{\cal C}^{(2)}_n=0$ for odd values of $n$. We take this opportunity to correct a typographical error present in our previous publications: the third line of Eq. (B2) in \cite{2024MNRAS.535..588I} and Eq. (A7) in \cite{2021PhRvD.104d3528S} incorrectly listed the leading coefficient of $\bar{\cal C}^{(2)}_4(\bar \gamma)$ as $\bar\gamma/2$. As written above, the correct coefficient is $\bar\gamma$. Finally, the relevant four-point propagator ($p=3$) at the two-loop order is given by:

\ba
\bar \Gamma_X^{(3)}(k,z) &=& {\rm exp}\left(-\bar G_\delta^2\,\gamma \right)\sum_n \bar G_X\,\bar G_\delta^{n-1} \bar{\cal C}^{(3)}_n\,\,; \\
\bar{\cal C}^{(3)}_3(\bar \gamma)&=&\bar F_X^{(3)}(\vec p,\vec q, \vec k-\vec p-\vec q) + \bar{\cal O}^{(3)}_{X,3}, \\
\bar{\cal C}^{(3)}_n(\bar \gamma) &=& \bar{\cal O}^{(3)}_{X,n}.
\ea

The above formalism dictates that fully evaluating a fiducial cosmology requires computing the specific coefficients $\bar{\cal C}^{(1)}_3$, $\bar{\cal C}^{(1)}_5$, $\bar{\cal C}^{(2)}_2$, $\bar{\cal C}^{(2)}_4$, and $\bar{\cal C}^{(3)}_3$. Once obtained, these yield the full suite of regularised propagators needed for Eq.~\ref{eq:pk_RegPT}. Because these statistics directly map to RegPT terms \cite{2012PhRvD..86j3528T}, our RegPT emulator outputs can be cleanly converted into predictions for $\bar{\cal C}_n$.   To map our fiducial theoretical templates, $\bar{P}^{PT}_{XY}$, onto target cosmologies featuring varying scale-independent parameters, we apply scaling relations directly to the underlying inputs of Eq.~\ref{eq:pk_RegPT}. Assuming the shape remains consistent, the linear power spectrum trivially scales via the growth factors:
\begin{equation}P_{\text{lin}} = \left(\frac{G_\delta}{\bar{G}_\delta}\right)^2 \bar{P}_{\text{lin}}.\end{equation}
The multi-point propagators undergo a similarly structured, albeit slightly more involved, transformation:

\begin{equation}\label{point_prop_new_cosmo_1}\Gamma_X^{(m)} = {\rm exp}\left(-G_X^2\gamma \right)\sum_n G_X G_\delta^{n-1} \bar{\cal C}^{(m)}(\gamma),
\end{equation}

where the dispersion parameter $\gamma$ must be explicitly rescaled:
\begin{equation}
\label{point_prop_new_cosmo_2}
\gamma=\left(\frac{G_\delta}{\bar G_\delta} \right)^2 \bar \gamma.\end{equation}Recall that $\bar\gamma=k^2\bar \sigma^2_{\rm d}/2$, where $\bar \sigma_{\rm d}^2$ is the displacement field dispersion computed from the initial linear power spectrum. Note that Eq.~\ref{point_prop_new_cosmo_2} corrects a second typographical error from our past work; in Eq. B10 of \cite{2024MNRAS.535..588I} and Eq. 2.25 of \cite{2018JCAP...07..018S}, the parameter $\gamma$ was missing this requisite scaling factor $(G_\delta / \bar G_\delta)^2$. Through this mapping, the $\bar{\cal C}_n$ coefficients generated by our fiducial RegPT emulators seamlessly construct the full template $P^{PT}_{XY}$ for any new target cosmology.

\bibliographystyle{JHEP}
\bibliography{Emulator}
\end{document}